\documentclass[a4paper]{jpconf}
\usepackage{graphicx}
\newcommand {\afrac} [2] {\left(\frac{#1}{#2}\right)}
\newcommand {\bbrac} [1] {\left[#1\right]}
\newcommand {\erf}       {{\rm erf}}
\newcommand {\vesc}      {v_{\rm esc}}
\newcommand {\xH}        {{\bf X}_{\rm H}}
\def \lsim {\:\raisebox{-0.7 ex}{$\stackrel{\textstyle<}{\sim}$}\:}
\def \gsim {\:\raisebox{-0.7 ex}{$\stackrel{\textstyle>}{\sim}$}\:}
\newcommand{\Includegraphics}  [2] {\includegraphics [width = 3.9 cm] {#1#2}}
\newcommand{\IncludegraphicsS} [2]
{%
\begin{center}%
 \Includegraphics {#1} {#2}
\end{center}
}
\newcommand{\IncludegraphicsD} [3]
{%
\begin{center}%
 \Includegraphics {#1} {#2}\hspace{2 cm}%
 \Includegraphics {#1} {#3}%
\end{center}
}
\newcommand{\IncludegraphicsQ} [5]
{%
\begin{center}%
 \Includegraphics {#1} {#2}%
 \Includegraphics {#1} {#3}%
 \Includegraphics {#1} {#4}%
 \Includegraphics {#1} {#5}%
\end{center}
}
\begin{document}
\title{Annual and diurnal modulations of                 \\ \vspace{-0.15cm}
       the angular distribution of the 3-D WIMP velocity \\ \vspace{-0.15cm}
       observed at an underground laboratory}
\author{Chung-Lin Shan}
\address{\it%
         Physics Division,
         National Center for Theoretical Sciences \\
         No.~101, Sec.~2, Kuang-Fu Road,
         Hsinchu City 30013, Taiwan, R.O.C.}
\ead{clshan@phys.nthu.edu.tw}
\begin{abstract}
 In this article,
 I present
 our (Monte Carlo) simulation results of
 the angular distribution of
 the 3-dimensional WIMP velocity,
 in particular,
 a possible ``annual'' modulation
 and the diurnal modulation
 proposed in literature,
 for different underground laboratories.
\end{abstract}
\section{Introduction}

 Directional direct Dark Matter detection experiments
 aim to identify
 the diurnal modulation of nuclear scattering signals
 induced by halo Weakly Interacting Massive Particles (WIMPs)
 coming (mainly) from the CYGNUS constellation.
 There are two kinds of diurnal modulation:
\begin{itemize}
\setlength{\itemsep}{0.025 ex}
\item
 Directionality:
 the diurnal modulation of the (main) incident direction of WIMP events;
 for a laboratory
 located in the Northern Hemisphere
 in Summer,
 the most WIMP events should come
 from the zenith around the midnight
 and from the north around the noon.
\item
 Flux shielding:
 the diurnal modulation of the number (scattering rate) of WIMP events;
 for a laboratory
 located in the Southern Hemisphere
 in Winter,
 the WIMP flux could be reduced in the day.
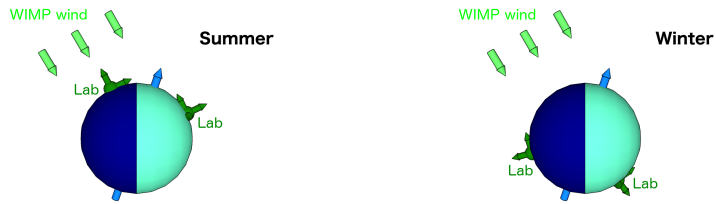
\IncludegraphicsD
 {skp-directional-1293-}
 {045N-summer}
 {035S-winter}
\end{itemize}
\subsection{Event generation in the Galactic coordinate system}

 For generating 3-D WIMP velocities
 in the Galactic coordinate system,
 we consider
 the simplest model of
 an isothermal, spherical and isotropic Dark Matter halo.
\begin{itemize}
\setlength{\itemsep}{0.025 ex}
\item
 The radial velocity distribution
 is the simple Maxwellian velocity distribution:
\begin{equation}
     f_{1, {\rm Gau}}(v)
  =  \bbrac{  \afrac{\sqrt{\pi}}{4} \erf\afrac{\vesc}{v_0}
            - \afrac{\vesc}{2 v_0}  e^{-\vesc^2 / v_0^2}   }^{-1}
     \afrac{v^2}{v_0^3}
     e^{-v^2 / v_0^2}
\,,
\label{eqn:f1v_Gau_vesc}
\end{equation}
 for $v \le \vesc$,
 and $f_{1, {\rm Gau}}(v > \vesc) = 0$.
\item
 The angular velocity distribution
 is simply isotropic:
\begin{equation}
     f_{\phi, {\rm G}}(\phi)
  =  1
\,,
     ~~~~ ~~~~ ~~ 
     \phi \in (-\pi,~\pi]
\,,
     ~~~~ ~       
\label{eqn:f1v_phi_G}
\end{equation}
 and
\begin{equation}
     f_{\theta, {\rm G}}(\theta)
  =  1
\,,
     ~~~~ ~~~~ ~~ 
    \theta \in [-\pi / 2,~\pi / 2]
\,.
\end{equation}
\item
 The measuring time
 of the recorded WIMP scattering events
 has also a constant probability:
\begin{equation}
     f_{t, {\rm G}}(t)
  =  1
\,,
     ~~~~ ~~~~ ~~\, 
     t \in [t_{\rm start},~t_{\rm end}]
\,.
     ~\,
\label{eqn:f1v_t_G}
\end{equation}
\end{itemize}

 Below is
 the angular WIMP velocity distribution
 generated with
 500 total events on average in one experiment
 in one entire year
 (i.e.,
  $[t_{\rm start},~t_{\rm end}] = [0, 365~{\rm day}]$),
 binned into 12 $\times$ 12 bins
 in the longitude and the latitude directions,
 respectively.
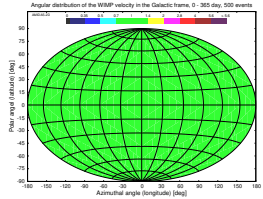
\IncludegraphicsS
 {N_}
 {phi_theta-G-500-00000}
 The horizontal color bar on the top of the plot
 indicates
 the mean value of the recorded event number
 (averaged over all 5,000 simulated experiments)
 in each angular bin
 in unit of the all--sky average value
 (500 events / 144 bins $\cong$ 3.47 events/bin here).

\section{Angular distribution of the 3-D WIMP velocity}

 In this section,
 I present the angular distribution patterns
 of the 3-dimensional WIMP velocity
 transformed to the laboratory--independent
 Equatorial coordinate system
 as well as
 the laboratory--dependent
 horizontal and laboratory coordinate systems
 (see Ref.~\cite{DMDDD-Nv}
  for their definitions).
 500 total events on average in one experiment
 in each (4-hour daily shift of each) 60-day observation period
 have been generated
 and
 5,000 experiments (for one laboratory)
 have been simulated%
\footnote{
 Note that,
 firstly,
 the annual modulations
 of the angular WIMP velocity distribution
 have been demonstrated
 for the advanced seasons
 centered on the dates of
 the February 19th (49.49 day),
 the May 21st (140.74 day),
 the August 20th (231.99 day),
 and
 the November 19th (323.24 day),
 respectively.
 On the other hand,
 for the diurnal modulations,
 four observation intervals of 4 hours
 at the central times of
  0 o'clock,
  6 o'clock,
 12 o'clock, and
 18 o'clock
 on the central dates of
 25.16 (= 390.16) day and 207.66 day
 have been considered.
 See Ref.~\cite{DMDDD-Nv}
 for details.
}.
\subsection{Annual modulation in the Equatorial coordinate system}

 Considering the very low event rate
 of one direct Dark Matter detector,
 especially
 a direction--sensitive TPC detector,
 we can,
 as the first step
 for identifying the directionality
 of incident WIMPs,
 combine WIMP events
 observed at different laboratories
 in the laboratory--independent
 Equatorial coordinate system:
 \IncludegraphicsQ
  {N_phi_theta-Eq-500-}
  {04949}
  {14074}
  {23199}
  {32324}
 Here
 the dark--green star
 (in all four plots)
 indicates
 the theoretical direction of the WIMP wind
 in the Equatorial coordinate system
 \cite{Bandyopadhyay10}:
 42.00$^{\circ}$S, 50.70$^{\circ}$W,
 while
 the blue--yellow point
 (in each plot)
 indicates
 the opposite direction of
 the Earth's relative velocity to the Dark Matter halo
 on the central date of {\em each} observation period
 (see Ref.~\cite{DMDDD-Nv}
  for the detailed calculations).

 It can be found that,
 firstly,
 the average event numbers
 from the center
 to the southwest part
 could be at least 11.4 times
 or even 16 times larger than
 the most part of the sky
 ($\gsim \, 13.9$ (19.4) events/bin
  against $\lsim \, 1.2$ events/bin
  among 500 total events).
 This would hence be a clear identification
 of the anisotropy of the main direction of incident WIMPs.
 Secondly,
 the distribution patterns in four plots variate slightly
 and
 this variation follows indeed
 the circular clockwise movement of the blue--yellow point.
 This would be,
 besides the pure ``directionality'' of the WIMP wind,
 a second (important) characteristic
 for identifying directional WIMP signals
 and discriminating from any (unexpected) backgrounds
 from some specified incoming directions.

\subsection{Annual modulation in the horizontal coordinate system}

 Once more and more WIMP events
 can be recorded
 in different laboratories,
 we can
 consider to demonstrate
 the angular distribution patterns
 of the 3-D WIMP velocity
 observed in the horizontal coordinate system
 of the laboratory of interest.
 Below are
 the angular WIMP velocity distribution
 observed in the horizontal coordinate systems
 of the Kamioka
 (36.43$^{\circ}$N, 137.31$^{\circ}$E)
 and the SUPL
 (37.07$^{\circ}$S, 142.81$^{\circ}$E)
 laboratories,
 respectively.
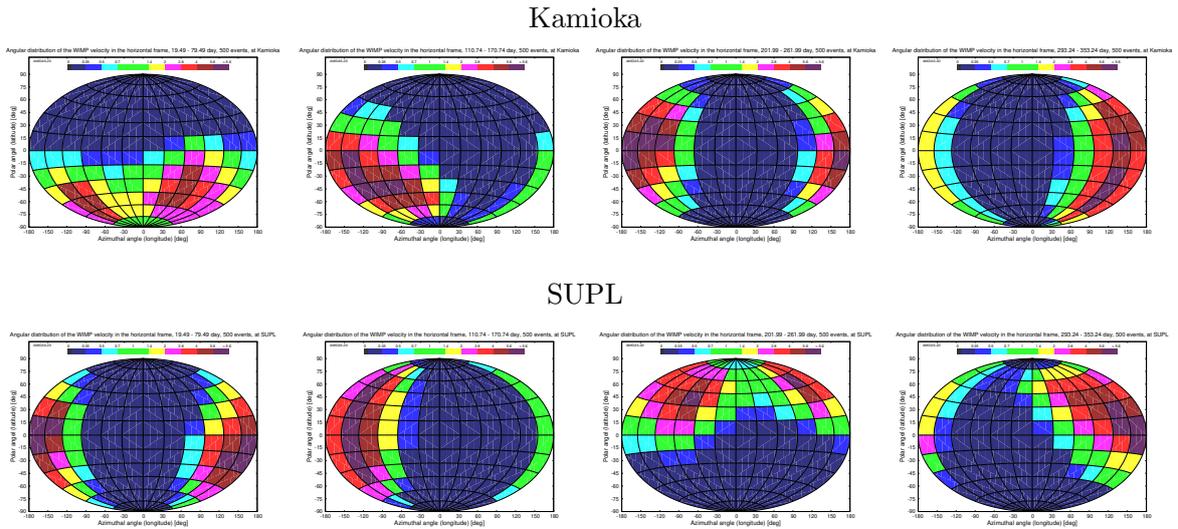
\def \LabName {Kamioka}
\begin{center}
 \LabName\
\end{center}
\vspace{-0.4 cm}
 \IncludegraphicsQ
  {N_phi_theta-H-500-}
  {04949-\LabName}
  {14074-\LabName}
  {23199-\LabName}
  {32324-\LabName}
\def \LabName {SUPL}
\begin{center}
 \LabName\
\end{center}
\vspace{-0.4 cm}
 \IncludegraphicsQ
  {N_phi_theta-H-500-}
  {04949-\LabName}
  {14074-\LabName}
  {23199-\LabName}
  {32324-\LabName}

 While
 SUPL is so far
 the unique functionable underground laboratory
 in the Southern Hemisphere,
 the Kamioka laboratory
 is located at the almost--symmetric point
 with respect to the Equatorial plane.
 This means that,
 with a 12-hour time difference in the same day
 or 
 a 6-month difference in a year,
 the horizontal coordinate systems
 of two laboratories
 should have a common $\xH$--axis
 (pointing towards north)
 and
 the difference between two frames
 is a 180$^{\circ}$--rotation
 around the common $\xH$--axis.
 Therefore,
 their angular distribution patterns
 show a 6-month difference
 with a 180$^{\circ}$--rotated symmetry
 around the center.

 Another interesting pair is
 the ANDES
 (30.19$^{\circ}$S, 69.82$^{\circ}$W)
 and the CJPL
 (28.15$^{\circ}$N, 101.71$^{\circ}$E)
 laboratories.
 They are at the almost opposite direction
 with respect to the Earth's center
 and thus
 the angular distribution patterns
 observed in their horizontal coordinate systems
 in the {\em same} season
 should have a 180$^{\circ}$--rotation difference
 around the center:
\newpage
\def \LabName {CJPL}
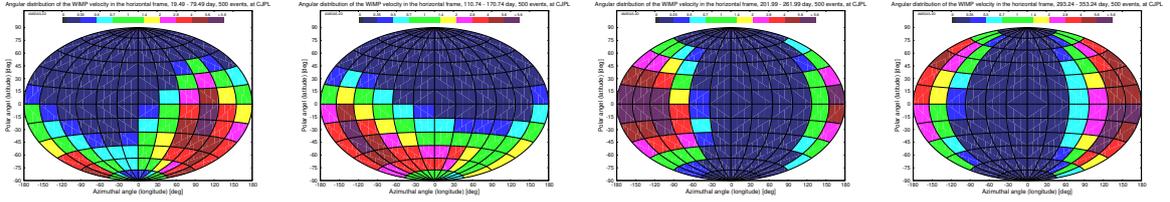
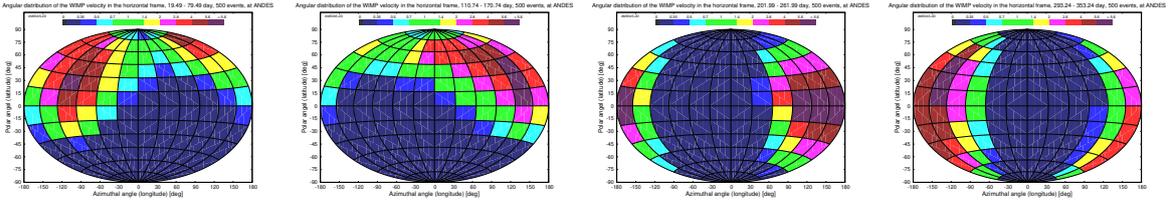
\begin{center}
 \LabName\
\end{center}
\vspace{-0.4 cm}
 \IncludegraphicsQ
  {N_phi_theta-H-500-}
  {04949-\LabName}
  {14074-\LabName}
  {23199-\LabName}
  {32324-\LabName}
\def \LabName {ANDES}
\begin{center}
 \LabName\
\end{center}
\vspace{-0.4 cm}
 \IncludegraphicsQ
  {N_phi_theta-H-500-}
  {04949-\LabName}
  {14074-\LabName}
  {23199-\LabName}
  {32324-\LabName}
 Moreover,
 the angular distribution pattern
 observed in the horizontal coordinate system
 of a North--American laboratory
 should be alike as
 that observed at an European laboratory
 in the {\em previous} season%
\footnote{
 Location information
 of the underground laboratories
 in North--America and Europe:
\\
 (1)
 DUSEL
 (44.35$^{\circ}$N, 103.75$^{\circ}$W),
 SNOLAB
 (46.47$^{\circ}$N,  81.19$^{\circ}$W),
 Soudan
 (47.82$^{\circ}$N, 92.24$^{\circ}$W);
\\
 (2)
 Boulby
 (54.55$^{\circ}$N,  0.82$^{\circ}$W),
 Callio
 (63.66$^{\circ}$N, 26.04$^{\circ}$E),
 LNGS
 (42.45$^{\circ}$N, 13.58$^{\circ}$E),
 LSBB
 (43.93$^{\circ}$N,  5.49$^{\circ}$E),
 LSC
 (42.81$^{\circ}$N,  0.56$^{\circ}$W),
 LSM
 (45.14$^{\circ}$N,  6.70$^{\circ}$E).
}:
\def \LabName {LNGS}
\begin{center}
 \LabName\
\end{center}
\vspace{-0.4 cm}
 \IncludegraphicsQ
  {N_phi_theta-H-500-}
  {04949-\LabName}
  {14074-\LabName}
  {23199-\LabName}
  {32324-\LabName}
\def \LabName {SNOLAB}
\begin{center}
 \LabName\
\end{center}
\vspace{-0.4 cm}
 \IncludegraphicsQ
  {N_phi_theta-H-500-}
  {04949-\LabName}
  {14074-\LabName}
  {23199-\LabName}
  {32324-\LabName}
\subsection{Diurnal modulation in the laboratory coordinate system}

 Finally,
 I present
 the diurnal modulation of
 the angular WIMP velocity distributions
 observed in the laboratory coordinate system
 of the Kamioka laboratory
 around 25.16 day and 207.66 day,
 respectively.
 From left to right,
 the simulated 4-hour event--measuring time
 are around
 midnight,
 morning,
 noon,
 and evening,
 respectively.
\def \LabName {Kamioka}
\begin{center}
 \LabName\ (25.16 day)
\end{center}
\vspace{-0.5 cm}
 \IncludegraphicsQ
  {N_phi_theta-Lab-500-39016-}
  {00-\LabName}
  {06-\LabName}
  {12-\LabName}
  {18-\LabName}
\vspace{-0.5 cm}
\begin{center}
 \LabName\ (207.66 day)
\end{center}
\vspace{-0.5 cm}
 \IncludegraphicsQ
  {N_phi_theta-Lab-500-20766-}
  {00-\LabName}
  {06-\LabName}
  {12-\LabName}
  {18-\LabName}
 It can be found firstly that,
 by comparing these two sets of plots
 in two observation periods
 with a half--year time difference
 with each other,
 the angular distribution patterns
 show indeed a 12-hour shift.
 Moreover,
 the variations of the diurnal modulation of
 the angular distribution patterns
 in both periods
 look similar to the annual modulation
 shown before.

 As a comparison,
 below
 I show also
 the diurnal modulation of
 the angular distribution patterns
 observed at the SUPL laboratory
 in two observation periods.
\def \LabName {SUPL}
\begin{center}
 \LabName\ (25.16 day)
\end{center}
\vspace{-0.5 cm}
 \IncludegraphicsQ
  {N_phi_theta-Lab-500-39016-}
  {00-\LabName}
  {06-\LabName}
  {12-\LabName}
  {18-\LabName}
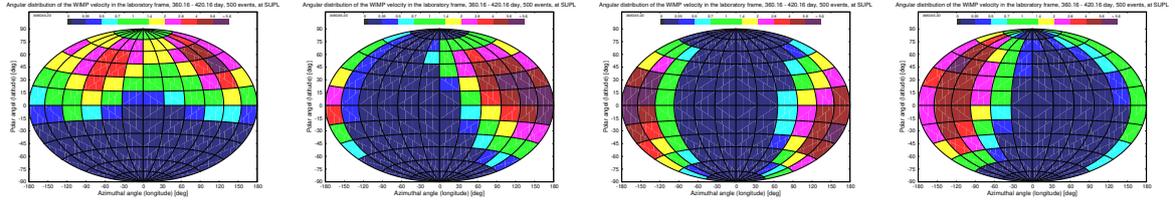
\vspace{-0.5 cm}
\begin{center}
 \LabName\ (207.66 day)
\end{center}
\vspace{-0.5 cm}
 \IncludegraphicsQ
  {N_phi_theta-Lab-500-20766-}
  {00-\LabName}
  {06-\LabName}
  {12-\LabName}
  {18-\LabName}
\section{Conclusions}

 In this article,
 I have demonstrated
 (the annual and diurnal modulations of)
 the directionality of
 the angular distribution
 of the Monte Carlo--generated 3-D WIMP velocity.
 For the first step
 with only a few observed WIMP events,
 one can combine events
 observed at different laboratories
 in the location--independent Equatorial coordinate system
 to identify the (variation of the) directionality
 of the WIMP wind.

 Once more and more WIMP events
 can be recorded
 in different laboratories,
 we can then compare
 the angular WIMP velocity distribution patterns
 in the horizontal coordinate systems
 of different laboratories.
 Instead of the confirmation/comparison of
 their diurnal modulations
 requiring a large number of recorded events,
 the annual modulation of
 the angular distribution pattern
 could be identified with less than 20\% WIMP events.

\section*{References}
\end{document}